\documentclass[letterpaper, 10 pt, conference]{ieeeconf}

\usepackage{graphicx} 
\usepackage{subfigure}
\usepackage{graphics} 
\usepackage{epsfig} 
\usepackage{times} 
\usepackage{amsmath} 
\usepackage{amssymb}  
\usepackage{url}
\newtheorem{theorem}{Theorem}

\newtheorem{algorithm}[theorem]{Algorithm}

\usepackage{cite}

\usepackage{color}
\usepackage{algorithm}
\usepackage[algo2e,ruled,linesnumbered]{algorithm2e}
\renewcommand{\KwIn}{$\textbf{Given:\quad}$}
\renewcommand{\KwOut}{$\textbf{Goal:\quad}$}

{ 

}

\newcommand{\bi}{\begin{itemize}}
	\newcommand{\ei}{\end{itemize}}
\newcommand{\bd}{\begin{displaymath}}
\newcommand{\ed}{\end{displaymath}}
\newcommand{\be}{\begin{eqnarray*}}
	\newcommand{\ee}{\end{eqnarray*}}

\newcommand{\dif}{\mathop{}\!\mathrm{d}}
\usepackage{caption2}

\IEEEoverridecommandlockouts
\begin{document}
	
	\title{Nonlinear Covariance Control via Differential Dynamic Programming}
	
	\author{Zeji Yi, Zhefeng Cao, Evangelos Theodorou, and Yongxin Chen
		\thanks{Z. Yi is with  the School of Aerospace Engineering, Georgia Institute of Technology, Atlanta, GA also the School of Aerospace Engineering, Tsinghua University, Bejing, China; {yizj16@mails.tsinghua.edu.cn}}
		\thanks{Z. Cao is with  the School of Aerospace Engineering, Georgia Institute of Technology, Atlanta, GA also the School of Aerospace Engineering, Zhejiang University, Hangzhou, China; {zcao99@gatech.edu}}
		\thanks{The first two authors contributed equally to this work.}
		\thanks{E. Theodorou and Y.~Chen are with the
			School of Aerospace Engineering, Georgia Institute of Technology, Atlanta, GA; {evangelos.theodorou@gatech.edu, yongchen@gatech.edu}}
		\thanks{Supported by the
		NSF under Grant ECCS-1901599.}
	}

	\maketitle

	\begin{abstract}
	We consider covariance control problems for nonlinear stochastic systems. Our objective is to find an optimal control strategy to steer the state from an initial distribution to a terminal one with specified mean and covariance. This problem is considerably more complicated than previous studies on covariance control for linear systems. We leverage a widely used technique -- differential dynamic programming -- in nonlinear optimal control to achieve our goal. In particular, we adopt the stochastic differential dynamic programming framework to handle the stochastic dynamics. Additionally, to enforce the terminal statistical constraints, we construct a Lagrangian and apply a primal-dual type algorithm. Several examples are presented to demonstrate the effectiveness of our framework.
	\end{abstract}
	
	
	\section{Introduction}
	
	The theory of covariance control/assignment was first studied by Hotz and Skelton \cite{HotSke87} in the 80's and further developed in \cite{IwaSke92,XuSke92,GriSke97}. The original goal was to find an optimal control strategy for a linear time-invariant system to achieve some specified stationary state covariance. Recently, the covariance control theory was extended to a finite horizon control setting \cite{CheGeoPav15a,CheGeoPav15b,CheGeoPav15c,HalWen16,Bak16,CheGeoPav18} where the goal was to steer the state covariance of a continuous-time linear dynamic system from an initial value to a terminal one. This finite-horizon perspective was further extended to cases with discrete-time dynamics \cite{Bak16a,Bak18}, nonlinear dynamics \cite{RidOkaTsi19}, multiple systems \cite{CheGeoPav19} etc. 
	The basic idea of covariance control is to relax the hard constraints in classical optimal control with soft probabilistic constraints; the former is usually unrealistically strong due to the stochastic disturbance. This relaxation makes it suitable for a range of applications in the presence of large uncertainties. Indeed, the covariance control theory has been applied to  in aerospace \cite{ZhuGriSke95,RidTsi18}, robotics\cite{OkaTsi19} and sensing \cite{Kal02} etc.

Most previous works on covariance control focused on linear systems. The goal of this paper is to generalize this framework to nonlinear stochastic dynamics. This will greatly expand the application domain as most real-world systems in robotics, autonomy, etc cannot be described within the linear dynamics regime. It turns out that the methods developed for linear system covariance control are not applicable to nonlinear problems. For one thing, the mean and the covariance in the linear system have independent dynamics and can be controlled separately and independently \cite{CheGeoPav15a}; this is no longer the case in nonlinear problems. 

In this work, we develop an efficient algorithm for nonlinear covariance control based on differential dynamic programming (DDP) \cite{JacMay70}, or more precisely, stochastic differential dynamic programming (SDDP) \cite{TheTasTod10}. 
To ensure the terminal constraint on the state mean and covariance, we adopt a Lagrangian multiplier method \cite{Ber14}. A primal-dual method is used to update the primal and dual variables iteratively. More specifically, for the given Lagrangian multiplier, SDDP is executed first to obtain the nominal trajectory and control. The multiplier is then updated following gradient direction, which is computed through propagating the dynamics forward under optimal control. 

The nonlinear covariance control problem was recently studied in \cite{RidOkaTsi19} under different assumptions with a different method. In particular, the cost function used in \cite{RidOkaTsi19} is quadratic. In addition, \cite{RidOkaTsi19} assumes additive noise whose intensity is independent of state and control. In contrast, we consider more general cost functions and dynamics. A case of particular interest to us is multiplicative noise which is ubiquitous in robotics \cite{TodLi05}. 
On another topic, even though differential dynamic programming has been developed for a long time, the literature with terminal constraints is relatively scarce \cite{JacMay70,SunTheTsi14} and all of them are for deterministic dynamics. How to generalize the methods to stochastic dynamics was not clear. It turns out that for problems with terminal constraints, there is a significant difference between stochastic settings and deterministic settings, as can be seen later in Section \ref{sec:noncov}. 

The paper is organized as follows. In Section \ref{sec:back} we provide the background on covariance control, differential dynamic programming, and Lagrangian multiplier method. The main result and the algorithm are presented in Section \ref{sec:noncov}. We present two examples in Section \ref{sec:eg} to illustrate the framework.  This follows by a short concluding remark in Section \ref{sec:conclusion}.

	\section{Background}\label{sec:back}
	\subsection{Covariance Control}
Covariance steering/control \cite{CheGeoPav15a} is about controlling the state of the stochastic system from the initial random vector $x(0)$ at $t=0$ to a terminal one $x(T)$ at $t=T$ via a control input that minimizes a certainty cost; here $x(0)$ has mean $\mu_0$ and covariance $\Sigma_{0}=\mathbb{E}[(x(0)-\mu_0)(x(0)-\mu_0)^{\mathrm{T}}]$, similarly, $x(T)$ has mean $\mu_T$ and covariance $\Sigma_{T}=\mathbb{E}[(x(T)-\mu_T)(x(T)-\mu_T)^{\mathrm{T}}]$. 
	The motivation stems from classical linear quadratic control problems\cite{CheGeoPav15a,Bak16}, in which, the objective is to steer the state of the system to
	the desired one in a way that strikes a balance between keeping the deviations of the system's state tolerable on average while also using affordable control effort.
	
	Covariance control provides enormous benefits in the integration of modeling, and control problems\cite{HotSke87}. In a general control task, except for controller, identification and state estimation also use covariances as a measure of performance as well. Therefore, a theory which steer covariance allows fusing the entire class of problems (modeling and control) of concern in systems via the same measure of performance.
	
	As for the linear system, the mean and the covariance have independent dynamics. Open-loop control is used to control the mean and the covariance is controlled by close-loop state feedback gain. However, in nonlinear system or constrained cases, the mean and covariance are usually coupled\cite{RidOkaTsi19}.
	
	\subsection{Differential Dynamic Programming}
	Differential dynamic programming (DDP) \cite{JacMay70} is an iterative algorithm for nonlinear optimal control problem, which has high execution speed so that is widely adopted. Consider a system with discrete-time dynamics
	\begin{equation}
	x(i+1)=f(x,u,i),\quad x(0)=\bar{x}_0,
	\end{equation}
	and cost function
	\begin{equation}
	J=\ell_f(x_N)+\sum_{i=0}^{N-1}{}\ell(x,u,i),
	\end{equation}
	where $N$ is the final time step, $x\in \mathbb{R}^{n}$ is the state of the system, $u\in \mathbb{R}^{m}$ is the control input, $\ell$ is the running cost, and $\ell_f$ is the final cost. Then, we define the value function at time i is the optimal cost-to-go starting at $x(i)=x$
	\begin{equation}
	V(x,i)\triangleq{}\min_u\limits\ \left[\ell(x,u,i)+V(f(x,u,i),i+1)\right],
	\end{equation}
	where $V(f(x,u,i),i+1)$ is value function at time $i+1$.
	
	The algorithm begins with a nominal trajectory, which is sequence of states $(\bar{x}_0,\bar{x}_1,\cdots,\bar{x}_N)$ and corresponding controls $(\bar{u}_0,\bar{u}_1,\cdots,\bar{u}_{N-1})$, and then executes a backward pass and a forward pass at each iteration. In the backward pass, the algorithm expands value function to second-order around the nominal trajectory\footnote{A modification of DDP with only first-order approximation of the dynamics was developed in \cite{TodLi05} under the name iLQR/iLQG.}. In the forward pass, the nominal trajectory will be updated using the optimal control law from the backward pass. The process will be repeated until convergence. 
	\subsubsection{Backward pass}
	First let $ Q(\delta x,\delta u,i) $ be the change in value function based on the i-th nominal $ (\bar{x}_i,\bar{u}_i) $ pair
	\begin{equation}
	\begin{aligned}
	Q(\delta x,\delta u,i)={}& \ell(x+\delta x,u+\delta u,i)\\
	&+V(f(x+\delta x,u+\delta u),i+1),
	\end{aligned}
	\end{equation}
	then approximate the cost-to-go function as a quadratic function, i.e. expand Q to second-order around  $ (\bar{x},\bar{u}) $
	\begin{equation}
	\begin{aligned}
	Q(\delta x,\delta u,i)={}&Q(i)+Q_{x}^{\mathrm{T}}(i)\delta x+Q_{u}^{\mathrm{T}}(i)\delta u\\
	&+\frac{1}{2}\delta x^{\mathrm{T}}Q_{xx}(i)\delta x+\frac{1}{2}\delta u^{\mathrm{T}}Q_{uu}(i)\delta u\\
	&+\frac{1}{2}\delta u^{\mathrm{T}}Q_{ux}(i)\delta x+\frac{1}{2}\delta x^{\mathrm{T}}Q_{xu}(i)\delta u,
	\end{aligned}
	\end{equation}
	where
	\begin{equation*}
	\begin{aligned}
	Q_{x}(i)={}&\ell_{x}(i)+f_x^{\mathrm{T}}V_{x}(i+1),\\
	Q_{u}(i)={}&\ell_{u}(i)+f_u^{\mathrm{T}}V_{x}(i+1),\\
	Q_{xx}(i)={}&\ell_{xx}(i)+f_x^{\mathrm{T}}V_{xx}(i+1)f_x+\kappa V_{x}(i+1)\cdot f_{xx},\\
	Q_{uu}(i)={}&\ell_{uu}(i)+f_u^{\mathrm{T}}V_{xx}(i+1)f_u+\kappa V_{x}(i+1)\cdot f_{uu},\\
	Q_{ux}(i)={}&\ell_{ux}(i)+f_u^{\mathrm{T}}V_{xx}(i+1)f_x+\kappa V_{x}(i+1)\cdot f_{ux}.
	\end{aligned}
	\end{equation*}
When $\kappa=1$, 2nd order expansion of the dynamics is used \cite{JacMay70}; When $\kappa=0$, 1st order expansion of dynamics is used\cite{TodLi05}. The parameter $\kappa$ is introduced to unify the results derived from the first and second-order expansion of the dynamics.
	
The optimal control law  $ \delta u^* $  of minimizing the quadratic approximation with respect to  $ \delta u $ is
	\begin{equation}
	\begin{aligned}
	\delta u^*&={}\underset{\delta u}{\operatorname{argmin}}\ Q(\delta x,\delta u) \\
	&={}k+K\delta x,
	\end{aligned}
	\end{equation}
	where
	\begin{equation}
	k=-Q_{uu}^{-1}Q_{u}\quad K=-Q_{uu}^{-1}Q_{ux}.
	\end{equation}
	Substitute  $ \delta u^* $  into the expansion of $Q$, we get a quadratic model of $V$
	\begin{equation}
	\begin{aligned}
	&V_x(i)={}Q_{x}(i)-Q_{xu}(i)Q_{uu}^{-1}(i)Q_u(i),\\
	&V_{xx}(i)={}Q_{xx}(i)-Q_{xu}(i)Q_{uu}^{-1}(i)Q_{ux}(i).
	\end{aligned}
	\end{equation}
	
	The backward pass begins by initializing the value function with the terminal cost also value function $ V(N)=\ell_f(x_N) $ and its derivatives, then calculate the value at each time.
	
	\subsubsection{Forward pass}
	After backward pass, we update the nominal trajectory using the optimal feedback control law got from the previous backward pass with the given initial condition
	\begin{subequations}
	\begin{eqnarray}
	x(0)&=&\bar{x}_0,\\
	u(i)&=&u(i)+k(i)+K(i)(x(i)-\bar{x}_i),\\
	x(i+1)&=&f(x,u,i).
	\end{eqnarray}
	\end{subequations}

	Finally, the backward pass and forward pass will be repeated until convergence when $k$ tends to be zero vector.

	\subsection{Lagrange multiplier}
	The method of Lagrange multipliers is a strategy for finding the maxima or minima of a function with constraints\cite{Ber14}
	\begin{equation*}
	\begin{aligned}
	\min_x \limits&\quad f_0(x)\\
	s.t.&\quad f_i(x)=0\quad i=1,\cdots, m.
	\end{aligned}
	\end{equation*}
	The basic idea is to convert a constrained problem into a form such that the derivative test of an unconstrained problem can still be applied. \par
	The Lagrangian dual problem is obtained by forming the Lagrangian of a minimization problem by using Lagrange multipliers to add the constraints to the objective function and then solving for the primal variable values that minimize the original objective function
	\begin{equation}
	\begin{aligned}
	g(\lambda)={}\min_{x}\left[f_{0}(x)+\sum_{i=1}^{m} \lambda_{i} f_{i}(x)
	\right],
	\end{aligned}
	\end{equation}
	where  $f_0$  is an original problem, $g$ is the unconstrained problem, $\lambda$ represents the Lagrange multiplier. This solution gives the primal variables as functions of the Lagrange multipliers, which are called dual variables so that the new problem is to maximize the objective function with respect to the dual variables under the derived constraints on the dual variables.\par
	

    In the past decades, several general methods for designing approximation algorithms for tricky optimization problems have arisen\cite{GoeWilDav97}, including the primal-dual method. The primal dual-algorithm, which is a standard tool in the design of algorithms for combinatorial optimization problems, while not a good general purpose LP solution technique, is valuable because it is easy to customize for a particular problem. Any feasible solution to the dual system can be used to initiate the algorithm. Associated with the dual solution is a "restricted" primal problem that requires optimization. When the solution of the restricted primal problem has been accomplished, and improved solution to the dual system can be obtained. This, in turn, gives rise to a new restricted problem to be optimized. Optimizing iteratively and the optimum is obtained for both primal and dual systems. The method has been used to solve problems that can be modeled as linear programs\cite{MonAdl89}. The constrained primal problem in our case is solved from a dual side also.
	
	\section{Nonlinear covariance control}\label{sec:noncov}
	
	We consider a class of nonlinear stochastic optimal control problems \cite{Aas12} with cost
	\begin{equation}
	J= \mathbb{E}\left[\ell_f(x(T))+\int_{0}^{T} \ell(x, u) \dif t \right],
	\end{equation}
	subject to the stochastic system described by the following stochastic differential equation \cite{Van07}
	\begin{equation}
	\dif x=f(x, u) \dif t+F(x, u) \dif \omega,
	\end{equation}
	where  $ x \in \mathbb{R}^{n} $ is the state, $u \in \mathbb{R}^{m} $  is the control,  $\omega \in \mathbb{R}^{p} $  is standard Brownian motion noise. The dynamic functions are $ f : \mathbb{R}^{n} \times \mathbb{R}^{m}\rightarrow \mathbb{R}^{n} $  and  $ F : \mathbb{R}^{n} \times \mathbb{R}^{m}\rightarrow \mathbb{R}^{n \times p} $. The term  $ \ell_f(x(T)) $  is the terminal cost while  $ \ell(x(t), u)$  is the instantaneous cost rate. The system starts from the initial condition
	\begin{equation}
	\begin{aligned}
	&\mathbb{E}[x(0)]=\mu_0,\\ &\mathbb{E}[(x(0)-\mu_0)(x(0)-\mu_0)^\mathrm{T}] = \Sigma_0,
	\end{aligned}
	\end{equation}
	where $ \mu_0 \in \mathbb{R}^{n} $ is the initial mean and $ \Sigma_0 \in \mathbb{R}^{n \times n} $ is the initial  covariance. We impose statistical terminal constraint
	\begin{equation}
	\begin{aligned}
	&\mathbb{E}[x(T)]=\mu_T,\\ &\mathbb{E}[(x(T)-\mu_T)(x(T)-\mu_T)^\mathrm{T}] = \Sigma_T,
	\end{aligned}
	\end{equation}
	where $ \mu_T $  is the desired mean and  $ \Sigma_T $  is the desired covariance. The overall value function  $ V(x, t) $  is defined as the optimal expected cost accumulated over the time horizon starting from the initial state  $ x $ at $t$ under optimal control.\par
	We first rewrite the constraint to $ \mathbb{E}[x(T)]=\mu_T$ and $\mathbb{E}[x(T)x(T)^{\mathrm{T}}] = \mu_T \mu_T^{{\mathrm{T}}}+\Sigma_T $, then use Lagrangian multiplier to derive an unconstrained problem of minimizing
	\begin{equation}
	\begin{aligned}
	&J={}\mathbb{E}\Bigl[\int_{0}^{T} \ell(x,u) \dif t +\lambda^{\mathrm{T}} (x(T)-\mu_T)\\
	&+\mathrm{tr}(\gamma^{\mathrm{T}} (x(T) x(T)^{{\mathrm{T}}}-\mu_T \mu_T^{{\mathrm{T}}}-\Sigma_T))\Bigl],
	\end{aligned}
	\end{equation}
	where $ \lambda \in \mathbb{R}^{n \times 1} $  and  $ \gamma \in \mathbb{R}^{n \times n} $  are the Lagrange multiplier. Now the term  $ \lambda^{{\mathrm{T}}} (x(T)-\mu_T)+\mathrm{tr}(\gamma^{\mathrm{T}} (x x^{{\mathrm{T}}}-\mu_T \mu_T^{{\mathrm{T}}}-\Sigma_T)) $  is the terminal cost. \par 
	
	To solve the adjoint problem concerning  $ u,\lambda,\gamma $  we may take the Lagrange multiplier into consideration in DDP and solve their optimal together to get a second order convergence rate. Another way is to use the duality, first solve u for the minima  $ g(\lambda,\gamma) $  then update the multiplier to find the optimal value for the dual problem with gradient ascend. After that solve the optimal u again, back and forth. Due to the existence of the Brownian motion term in dynamics, the introduction of the second order term with respect to $ \lambda ,\gamma $  will recursively bring higher order terms of the value function. In particular, $Q_{\lambda\lambda}$ contains $V_{xx\lambda\lambda}$, because of the noise term in the expansion of $\delta x(t+\delta t)$.  As a result, second order convergence rate is not achievable. Therefore we recommend the second option. Extra attention should be paid that the value function must be convex at the ending point, therefore  $ \gamma $  should be positive definite in each iteration.\par
	\subsection{System Dynamics Linearization and Discretization}
	Given a nominal trajectory of states and controls we do expansion with respect to  $ x $  and  $ u $  up to the second order ($\kappa=1$) for differential dynamic programming or the first order ($\kappa=0$) for iLQG \cite{TodLi05}. Then discretize the system with  $ \delta t=t_{i+1}-t_{i} $  corresponding to a small interval to transit it from continuous to discrete time. All the derivations from here on consider first order terms with respect to time and up to second order terms with respect to state. The expansion expressed as
	\begin{equation}
	\begin{aligned}
	\delta x(t+\delta t)={}&A_{t} \delta x(t)+B_{t} \delta u(t)+\Gamma_{t} \boldsymbol{\xi}(t)\\
	&+\kappa\mathbf{O}_{d}(\delta x, \delta u, \boldsymbol{\xi}, \delta t),
	\end{aligned}
	\end{equation}
	where the random variable  $ \boldsymbol{\xi} \in \mathbb{R}^{p \times 1} $  is zero mean and Gaussian distributed with covariance $\sigma^{2} I_{p\times p} $ while the matrices  $ A_{t} \in \mathbb{R}^{n \times n}, B_{t} \in \mathbb{R}^{n \times m} $  and  $ \Gamma_{t} \in \mathbb{R}^{n \times p} $  are defined as\cite{TheTasTod10}
	\begin{equation}
	\begin{aligned}
	A_{t} &=I_{n \times n}+\nabla_{x} f(x, u) \delta t, \\ 
	B_{t} &=\nabla_{u} f(x, u) \delta t, \\ 
	\Gamma_{t} &=\left[\begin{array}{llll}{\Gamma^{(1)}} & {\Gamma^{(2)}} & {\ldots} & {\Gamma^{(m)}}\end{array}\right],
	\end{aligned}
	\end{equation}
	with $\Gamma^{(i)} \in \mathbb{R}^{n \times 1}$  defined  $ \Gamma_{t}=\Delta_{t}(\delta x, \delta u)+F(x, u) $ where each column vector of  $ \Delta_{t} $  is defined as  $ \Delta_{t}^{(i)}(\delta x, \delta u)=\nabla_{u} F_{c}^{(i)} \delta u(t)+\nabla_{x} F_{c}^{(i)} \delta x(t) $. Notice that  $ F_c^{(i)} $  represent the ith columns of  $ F $.
	Also  $ \mathbf{O}(\delta x, \delta u, d \omega) \in \mathbb{R}^{n \times 1} $  contains all the second order terms in the deviations in states, controls and noise. 
	
	As in original DDP, the derivation of SDDP requires the second order expansion of the action value function around a nominal trajectory  $ \bar{x} $. Substitution of the discretized dynamics in the second order value function expansion results in
	\begin{equation}
	\begin{aligned}
	&V\left(\bar{x}+\delta x,t+\delta t\right)=\\
	&V\left(\bar{x},t+\delta t\right)+V_{x}(\bar{x}+\delta x,t+\delta t)^{{\mathrm{T}}}\\
	&\left(A_{t} \delta x(t)+B_{t} \delta u(t)+\Gamma_{t} \boldsymbol{\xi}(t)+\kappa\mathbf{O}_{d}\right) \\ 
	&+\left(A_{t} \delta x(t)+B_{t} \delta u(t)+\Gamma_{t}\boldsymbol{\xi}(t)+\kappa\mathbf{O}_{d}\right)^{{\mathrm{T}}} \\ 
	&V_{x x}(\bar{x}+\delta x,t+\delta t)\\
	&\left(A_{t} \delta x(t)+B_{t} \delta u(t)+\Gamma_{t} \boldsymbol{\xi}(t)+\kappa\mathbf{O}_{d}\right).
	\end{aligned}
	\end{equation}
	
	After calculating the expectations of the terms with uncertainty and reshape the matrix, we recollect all the first order and second order expansion of the value function $V$ and get the following action value function 
	\begin{equation}\label{eq:DDP_Q}
	\begin{aligned}
	Q_{x}(i) &=\ell_{x}+A_{t} V_{x}(i+1)+\tilde{\mathcal{S}}, \\
	Q_{u}(i) &=\ell_{u}+A_{t} V_{x}(i+1)+\tilde{\mathcal{U}}, \\ 
	Q_{x x}(i) &=\ell_{x x}+A_{t}^{{\mathrm{T}}} V_{x x}(i+1) A_{t}+\kappa\mathcal{F}+\tilde{\mathcal{F}}+\tilde{\mathbf{M}}, \\
	Q_{x u}(i) &=\ell_{x u}+A_{t}^{{\mathrm{T}}} V_{x u}(i+1) B_{t}+\kappa\mathcal{L}+\tilde{\mathcal{L}}+\tilde{\mathbf{N}}, \\
	Q_{u u}(i) &=\ell_{u u}+B_{t}^{{\mathrm{T}}} V_{u u}(i+1) B_{t}+\kappa\mathcal{Z}+\tilde{\mathcal{Z}}+\tilde{\mathbf{G}}, \end{aligned}
	\end{equation}
	where the matrices  $ \mathcal{F} \in \mathbb{R}^{n \times n} $,  $ \mathcal{L} \in \mathbb{R}^{m \times n} $,  $ \mathcal{Z} \in \mathbb{R}^{m \times m} $,  $ \tilde{\mathcal{U}} \in \mathbb{R}^{m \times 1} $,  $ \tilde{\mathcal{S}} \in \mathbb{R}^{n \times 1} $,  $ \tilde{\mathcal{F}} \in \mathbb{R}^{n \times n} $, $ \tilde{\mathcal{L}} \in \mathbb{R}^{n \times m} $, $ \tilde{\mathcal{Z}} \in \mathbb{R}^{m \times m} $, $ \tilde{\mathbf{M}} \in \mathbb{R}^{n \times n} $,  $ \tilde{\mathbf{N}} \in \mathbb{R}^{n \times m} $  and  $ \tilde{\mathbf{G}} \in \mathbb{R}^{m \times m} $ are defined as in the SDDP paper\cite{TheTasTod10}.
	To find the optimal control policy, we compute the local variations in control  $ \delta u(i) $  that minimize the Q-function
	\begin{equation}
	\begin{aligned} 
	\delta u^{*} &=\underset{u}{\operatorname{argmin}} \ Q(\bar{x}+\delta x, \bar{u}+\delta u) \\ 
	&=-Q_{u u}^{-1}\left(Q_{u}+Q_{u x} \delta x\right). \end{aligned}
	\end{equation}
	
	We propagate backward in time iteratively to get the second-order local approximation of the value function. The controller is used to generate a locally optimal trajectory by propagating the dynamics forward in time. Substituting $ \delta u $ in second order expansion of action value function with the optimal control policy, we get the update law of value function at each time.
	\begin{equation}\label{eq:DDP_V}
	\begin{aligned}
	V(i)={}&V(i)-\frac{1}{2}Q_{u}(i) Q_{u u}^{-1}(i) Q_{u}(i), \\ V_{x}(i)={}&Q_{x}(i)-Q_{u}(i) Q_{u u}^{-1}(i) Q_{u x}(i), \\
	V_{x x}(i)={}&Q_{x x}(i)-Q_{x(i) u} Q_{u u}^{-1}(i) Q_{u x}(i).
	\end{aligned}
	\end{equation}
	\subsection{Multiplier Update}
	We continue with the control value we get from with fixed  $ \lambda $  and  $ \gamma $  and extract the control policy for each time step. Notice that what we get is not only a open loop control value  $ u_0 $  but also a closed loop policy  $ \delta u = k+K \delta x $. When we execute the SDDP method to convergence we will get  $ Q_u=0 $  at each time step. Therefore  $ \delta u = Q_{uu}^{-1}Q_{ux}\delta x $. When the algorithm converges,  $ \delta x=0 $  but the control policy is still valuable and provides the closed loop part which represented as  $ K=Q_{uu}^{-1} Q_{ux} $  here. The overall control policy can be expressed as
	\begin{equation}
	u(i)=K(i) (x(i)-\bar{x}_i)+\bar{u}_i,
	\end{equation}
	where $ \bar{x} $  is the trajectory which we propagate according to dynamics without noise as in SDDP.
	$ \bar{u} $  is what we solved for in SDDP. When convergence is reached for fixed  $ \lambda $  and  $ \gamma $, we compute  $ \mathbb{E}[x(T)] $  and  $ \mathbb{E}[x(T)x(T)^{\mathrm{T}}] $, and then update the multipliers with gradient ascend method\cite{GoeWilDav97}
	\begin{equation}\label{eq:Lagrangian}
	\begin{aligned}
	 V_{\lambda}(T)&= \mathbb{E}[x(T)]-\mu_T,\\
     V_{\gamma}(T)&=\mathbb{E}[x(T)x(T)^{\mathrm{T}}]-\mu_T\mu_T^\mathrm{T}-\Sigma_T.
	\end{aligned}
	\end{equation} 
	
	There are two ways to compute the above terms. In the first one, we sample the noise, propagate the stochastic dynamics and calculate the expectation in a statistic way. The feedback law is employed with each step, where  $ x $  in the control law is the actual state we propagate with real noise. The second method is to propagate the mean and covariance half analytically\cite{PanThe14}. The first way has better feasibility but will oscillate slightly even at the fixing point. Consider the samples as i.i.d. and the covariance of  $ \mathbb{E}[x] $  and  $ \mathbb{E}[x x^{\mathrm T}] $  are of  $ {1}/{\sqrt{n}} $  order of magnitude. The required number of samples is proportional to the square of resolution. For higher resolution, increasing the number of samples is necessary, which will also decrease the randomness in gradient descent, therefore decrease the number of steps to convergence. However, there is always a trade-off between computation speed and precision, which we recommend a combination of several hundred samples and a ten percent resolution. As for the second method, the problem is that for some highly nonlinear cases it may be hard to propagate analytically and the approximation of each state to be Gaussian may have an unneglectable offset for some certain cases especially after many steps. The proposed algorithm is summarized in Algorithm \ref{alg:SDDP}.
	\begin{algorithm}[h]
		\caption{Covariance Control with SDDP}\label{alg:SDDP}%
		\KwIn{Initial state mean $\mu_0$ and covariance $\Sigma_0$, total time step $N$, control sequence $\bar{u}$, initial Lagrange multiplier $\lambda_0$ and $\gamma_0$, and target mean $\mu_T$ and covariance $\Sigma_T$}. Set $\kappa=1$ for the case of 2nd order dynamics expansion and  $\kappa=0$ for 1st order expansion of dynamics. 
		\\
		\KwOut{Optimal control sequence $u^*$, corresponding state trajectory $x^*$ and local state feedback control strategy $K_i,i=1,\cdots,N-1$}
		\\
		\\
		Get initial trajectory $\bar{x}$ by integrating the dynamics forward with $ \bar{x}_0$ and $\bar{u}$;\\
		\While{not convergence,}{
			Differentiate value function at final time, get $V_x(N)$, $V_{xx}(N)$;\\
			\For{$i=N-1,\cdots,1$}{
				Compute the value of $Q_x$, $Q_u$, $Q_{xx}$, $Q_{xu}$, $Q_{ux}$, $Q_{uu}$ at time i according to \eqref{eq:DDP_Q};\\
				Compute the value of $V_x$, $V_{xx}$ at time i according to \eqref{eq:DDP_V};}
			\For{$i=1,\cdots,N-1$}{
				Update control sequence $\bar{u}=\bar{u}+\delta u^*$ with control policy $\delta u^*=k+K\delta x$ at time $i$;\\
				Update state trajectory $\bar{x}$ at time $i$  without noise;}
			\If{SDDP not convergence,}{
				continue;
			}
			Sample trajectories from initial state mean $\mu_0$ and covariance $\Sigma_0$ using forward dynamics with noise;\\
			Get statistic final state's mean $\mu(N)$ and covariance $\Sigma(N)$;\\
			Compute the gradients of Lagrange multiplier $\lambda$ and $\gamma$ according to \eqref{eq:Lagrangian} and update Lagrange multiplier $\lambda=\lambda+\eta_1 V_{\lambda}(T)$ and $\gamma=\gamma+\eta_2 V_{\gamma}(T)$, where $\eta_1\in [0,1]$ and $\eta_2\in [0,1]$ are search parameters ;
		}	
	\end{algorithm}

	\section{Numerical Examples}\label{sec:eg}
	In this section, we demonstrate the algorithm's performance with the specified terminal constraint on two different systems. We focus on the final state mean and covariance to meet the terminal constraint. 
	\subsection{One-Dimensional Dynamics}
	First, we consider one-dimensional stochastic nonlinear system of the form
	\begin{equation}
	\dif x=\cos (x) \dif t+u \dif t+x^{2} \dif \omega.
	\end{equation}
	Our goal is to manipulate the system with a state dependent noise to reach a final state $x(T)$ with zero mean and covariance $0.03$. The initial state has mean $0$ and covariance $0.25$. The running cost is $\ell=\mathbb{E}\left[\int_{0}^{T} r u^{2} d \tau\right]$. This is augmented by terminal cost $\ell_f=\mathbb{E}\left[\lambda( x(T)-\mu_T)+\gamma (x(T)^2-\mu_T^2-\Sigma_T)\right]$ where $r=10^{-4}$ induced from Lagrangian multiplier. The time step is set to be $ 10 $ msec and time horizon is $ T=1 $ sec. We sample the noise and get several trajectories to calculate gradients $ V_\lambda(T) = \mathbb{E}[x(T)]-\mu_T $, and $ V_\gamma(T)={}\mathbb{E}[x(T)^2-\mu_T^2-\Sigma_T]$ statistically, while the number of sample is related to the distance to the optimal point. It's worth noting that $\gamma$ should be positive during the updating process to guarantee the convexity of the terminal cost. In the beginning, only a rough direction is needed for those $\lambda$ and $\gamma$ who are distant from the optimal point, i.e. $V_\lambda$ and $V_\gamma$ are rather large, thus the number of samples can be small, in our case we choose 80. While for those $\lambda$ and $\gamma$ which are rather close to the optimum, i.e. $V_\lambda$ and $V_\gamma$ are rather small, the gradient ascent direction and length should be relatively precise. So that the next step will get towards the optimum instead of drifting around because of the uncertainty in the calculation of mean and covariance.

\begin{figure}[htbp]
		\includegraphics[width=1\columnwidth]{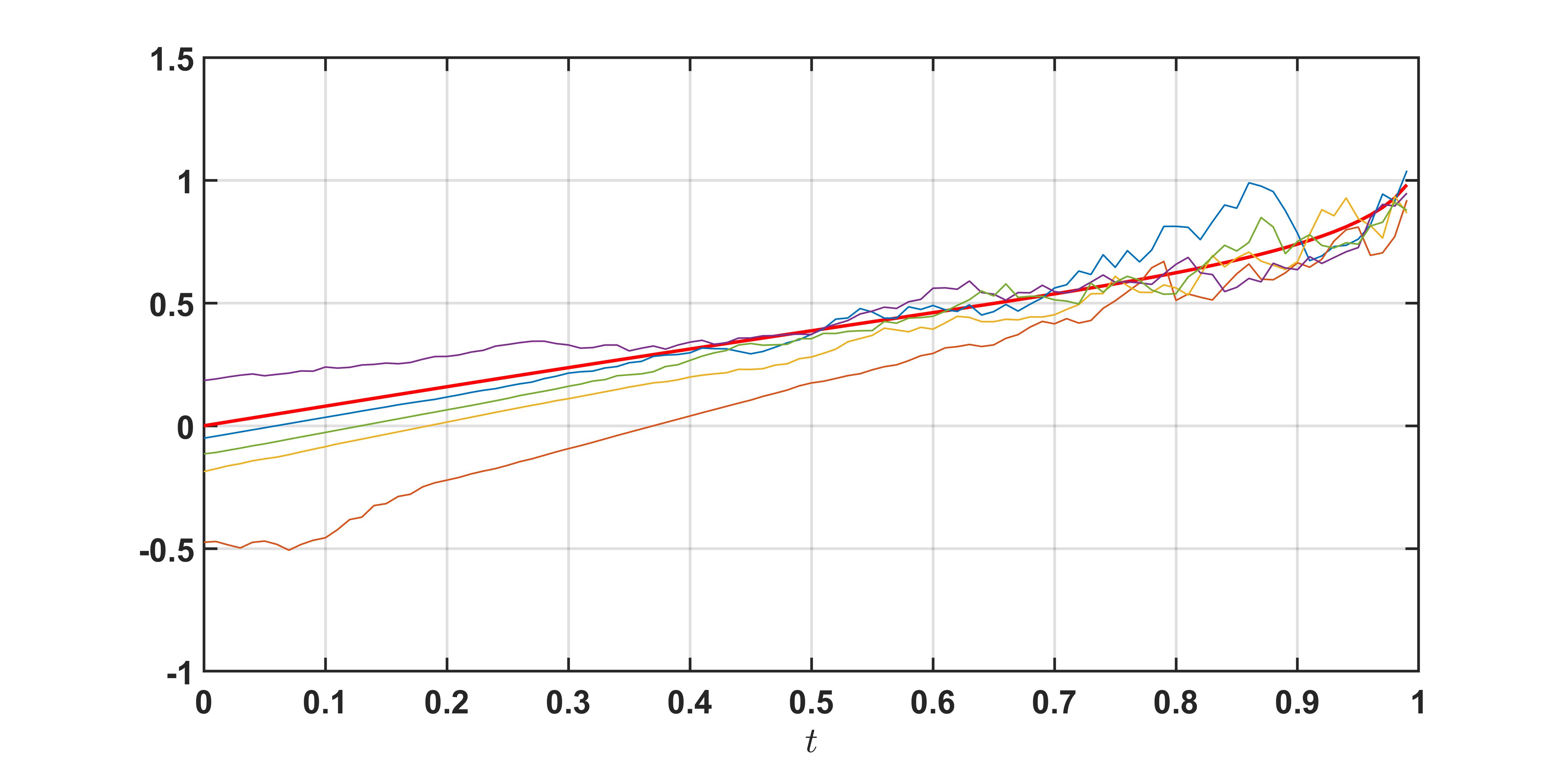}
		\caption{The nominal trajectory is marked with the red line. The colored lines are several sampled trajectories.}
		\label{fig:coss}
	\end{figure}

\begin{figure}[htbp]
		\includegraphics[width=1\columnwidth]{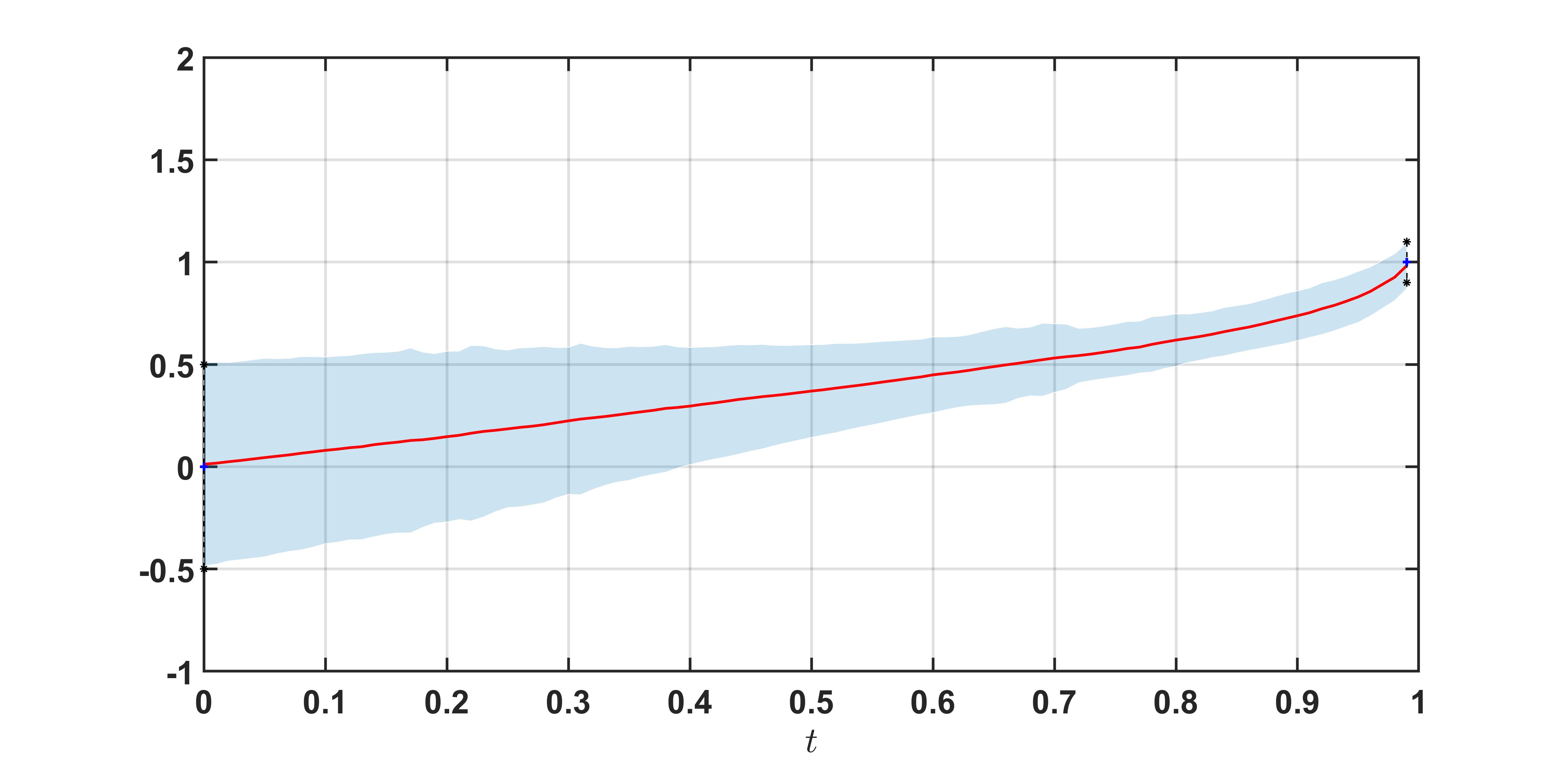} 
		\caption{The whole process's mean is overlapped with the nominal trajectory as in the red line. The shadow part illustrates state covariance.}
		\label{fig:cosc}
	\end{figure}

	From the experiment, the resolution of the final state is about 0.01 for both mean and covariance with 800 samples around the neighborhood. Also the covariance becomes much smaller under the feedback control while starting with $\Sigma_0 = 0.25$ and ending with $\Sigma_T = 0.01$. As a comparison, for the uncontrolled case, the mean and covariance nearly remain the same. Fig. \ref{fig:coss} display the nominal state trajectory under feedback control with some sampled trajectories. Fig. \ref{fig:cosc} illustrates the nominal trajectory's shift with covariance which corresponds to our controlled system well.
    
	\subsection{Simple Inverted Pendulum}
\begin{figure}[htbp]   
	\includegraphics[width=1\columnwidth]{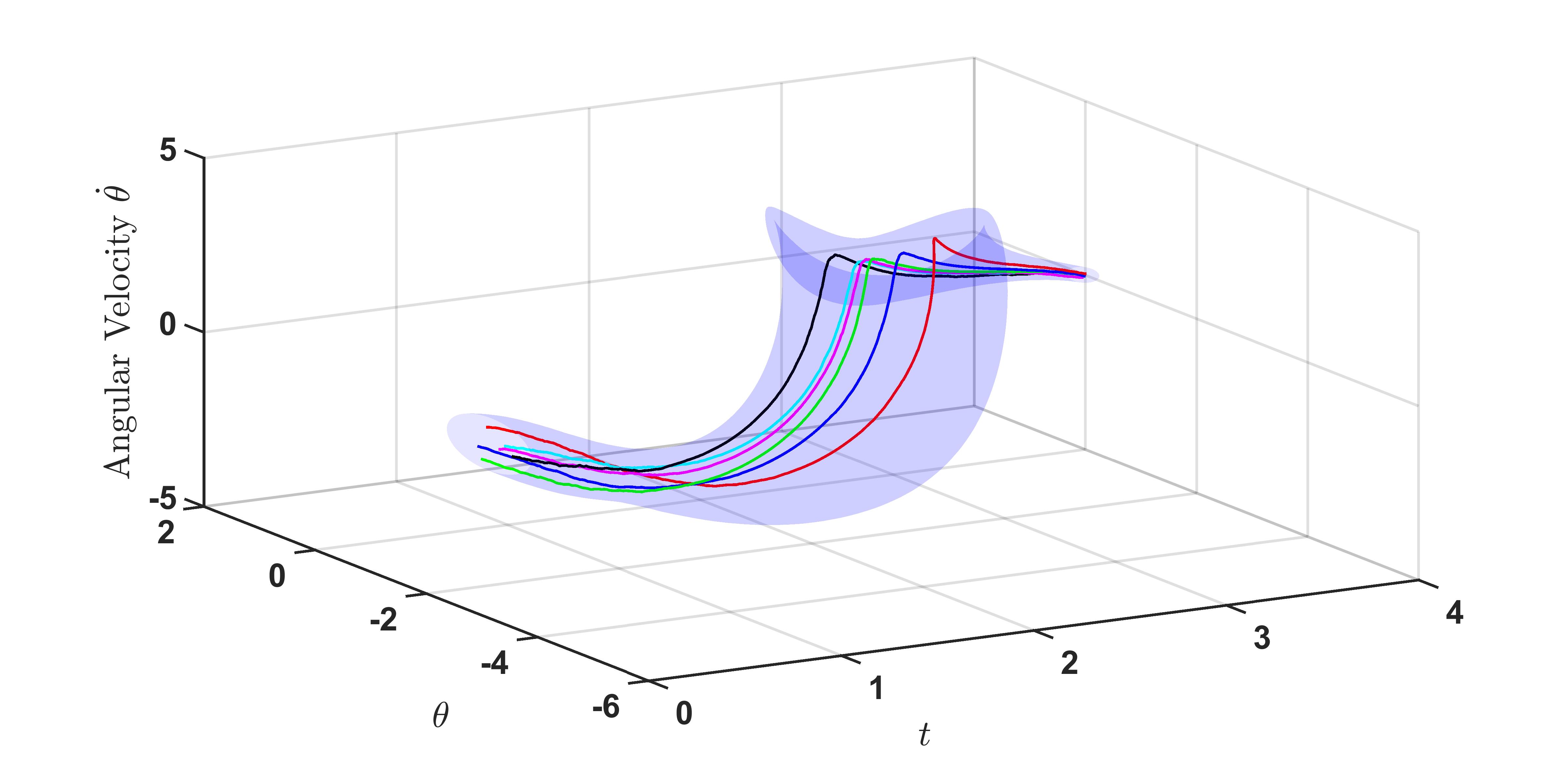} 
	\caption{Colored lines are the sampled trajectories and the blue shadowed part indicates their covariance (3-$\sigma$ region.}
	\label{fig:pencov}     
	\end{figure}

\begin{figure}[htbp]  
		\includegraphics[width=1\columnwidth]{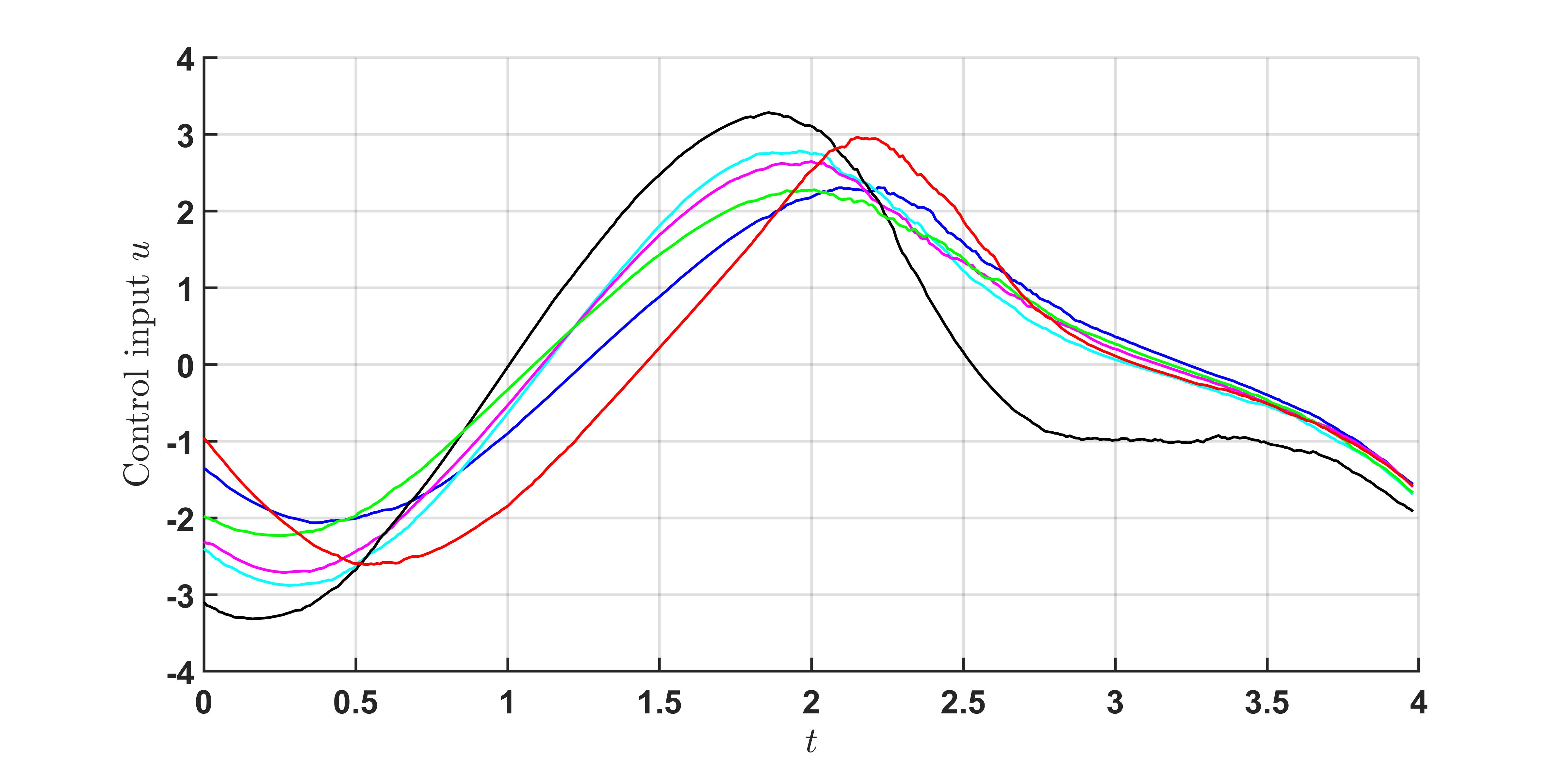}  
		\caption{Control trajectories of the sampled ones are displayed correspondingly.} 
		\label{fig:penc}   
	\end{figure}

\begin{figure}[htbp]
		\includegraphics[width=1\columnwidth]{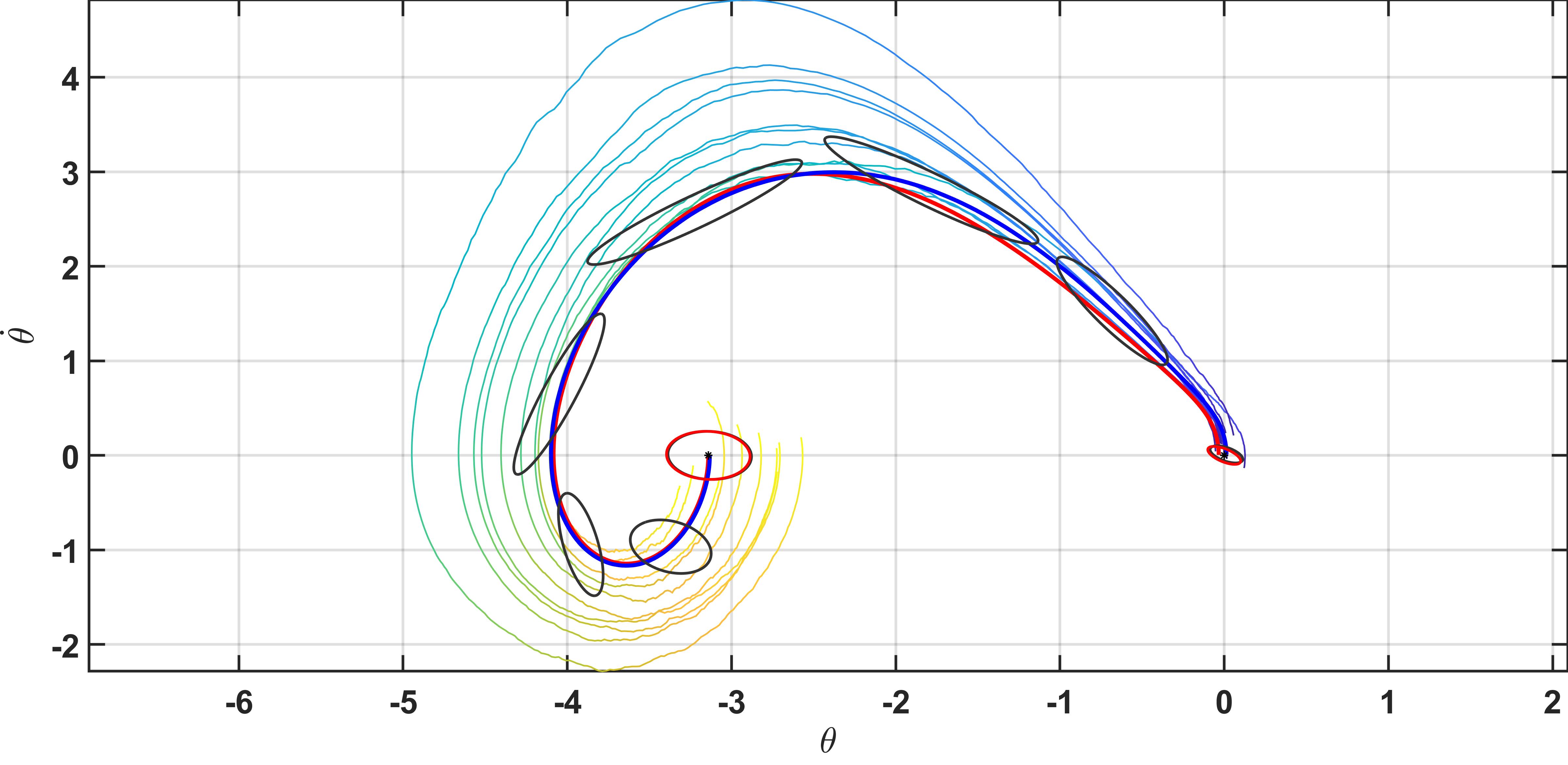}
		\caption{The Red line indicates the nominal trajectories. The Blue line indicates the mean of samples. Several sampled trajectories are presented with colored lines whose variation in color symbolizes the passage of time. The one-sigma ellipses of some selected time points' are presented as well.}
		\label{fig:penp} 
	\end{figure}

	The second model we study is an inverted pendulum, with constraints on both final state mean and covariance. The stochastic nonlinear dynamics is 
	\begin{subequations}
		\begin{eqnarray}   
			\dif x_1&=&x_2 \dif t,\\
			\dif x_2&=&4sin(x_1)\dif t+u\dif t+\alpha u\dif \omega,
		\end{eqnarray}
	\end{subequations}
	where the state variables are $x_1=\theta, x_2=\dot{\theta}$, the measurement parameter of noise $ \alpha =0.04$, the time step is $ 10 $ msec and time horizon is $ T=4 $ sec. The goal is to let the suspended pendulum swing up from initial condition (corresponding to a -180 deg angle) to an inverted position (corresponding to a 0 deg angle) and stay static at the final time $T$. The first step is to find the optimal trajectory and the corresponding feedback control strategy with the following cost function when the Lagrange multipliers are fixed
	\begin{equation}
	\begin{aligned}
	J=&\mathbb{E}\{\lambda^{\mathrm{T}}(x(T)-\mu_T)+x(T)^{\mathrm{T}}\gamma x(T)-\mu_T^{\mathrm{T}}\gamma \mu_T-\Sigma_T\\
	&+0.01\int_{0}^{T}u(t)^2\dif t\}.
	\end{aligned}
	\end{equation}
	As in the previous example, we sample the noise and get some trajectories to calculate gradients. Now it's necessary to guarantee that the multiplier matrix $ \gamma $ should be positive define. In this case, we choose 800 for the number of samples. Fig. \ref{fig:pencov} demonstrates the variety of the mean and covariance with several sampled trajectories. Fig. \ref{fig:penc} displays the control sequences of the sampled trajectories in Fig. \ref{fig:pencov}. Fig. \ref{fig:penp} illustrates all nominal trajectory, sampled trajectories, mean and covariance variety in a phase graph.

	\section{Conclusion}\label{sec:conclusion}
	In this paper, we developed an algorithm for covariance control of nonlinear stochastic systems. Our method is based on stochastic differential dynamic programming \cite{TheTasTod10}. In order to achieve targeted state mean and covariance, SDDP is combined with the Lagrangian multiplier method to handle the terminal constraints. We remark that DDP with terminal constraints for deterministic systems doesn't apply in a stochastic setting. We tested our algorithm on severally examples and observed satisfying performance. The next step is to apply this algorithm for high-dimensional robotics path/trajectory planning problems. There are also several potential directions to improve the performance of our algorithm including working on belief space \cite{PanThe14}.

	\bibliographystyle{IEEEtran}
	\bibliography{refs}
	
\end{document}